\documentclass[10pt,preprint]{emulateapj}
\usepackage{amsmath}
\usepackage{lineno}
\usepackage{apjfonts}
\shorttitle{Asteroseismology of the DAV star L19-2}
\shortauthors{Y. H. Chen}

\begin{document}

  \slugcomment{Submitted to ApJ}

  \received{}
  \revised{}
  \accepted{}

  \title{Asteroseismology of the DAV star L19-2}

\author{Yanhui Chen\altaffilmark{1,2,3}}
\altaffiltext{1}{Institute of Astrophysics, Chuxiong Normal University, Chuxiong 675000, China; yanhuichen1987@126.com}
\altaffiltext{2}{School of Physics and Electronical Science, Chuxiong Normal University, Chuxiong 675000,China}
\altaffiltext{3}{Key Laboratory for the Structure and Evolution of Celestial Objects, Chinese Academy of Sciences, P.O. Box 110, Kunming 650011, China}

  \begin{abstract}

L19-2 is a DAV star, which has been intermittently observed from 1976 to 2013. Five independent pulsation modes of 350\,s, 192\,s, 143\,s, 118\,s, and 113\,s are identified. The five modes can be used to constrain the fitting models. The rates of period change can be obtained through the O-C method for modes of 192\,s and 113\,s, which can be used to study the evolution effect of DAV stars. Using the \texttt{WDEC} (2018 version), a large sample DAV star models are evolved. The theoretical modes are calculated and used to fit the observed modes. After fine model fittings, we obtain an optimal model with an absolute difference of $\Phi$ = 0.06\,s. By parameterizing the core oxygen profile, the \texttt{WDEC} (2018 version) procedure can greatly reduce the fitting error of asteroseismological model. According to our optimal model, the distance obtained through the model luminosity is only 1\% different from that reported by the Gaia Data Release 2. The L19-2 is a massive and hot DAV star with relatively a thick H atmosphere and a thick He layer. The stellar parameters and the rates of period change of our optimal model are a little modifications to that of the previous work. Our optimal model has a large central oxygen abundance. The central oxygen abundance is strongly correlated with the previous physical process of stellar evolution. A lot of asteroseismological work on white dwarfs have an opportunity to explore the progenitor stars.

  \end{abstract}

  \keywords{asteroseismology: individual (L19-2)-white dwarfs}


\section{Introduction}

Asteroseismology is a powerful tool to detect the internal structure of stars by using their oscillations as seismic waves. Handler (2013) reviewed the driving mechanisms, the basic idea of asteroseismology, and the properties of all kinds of variable and pulsating stars. In recent decades, the classes of pulsating variable stars have increased significantly, and our understanding of pulsating objects has greatly deepened. The study of asteroseismology can help us understand the internal structure of stars better and test and improve the theory of stellar structure and evolution. It is one of the research hot spots in the field of astrophysics.

White dwarfs (\texttt{WD}s) are the final evolutionary stage of 98\% of all stars in the universe (Winget \& Kepler 2008). \texttt{WD}s are extremely dense objects, which are composed of electron degenerate central core and ideal gas surface atmosphere. The thermonuclear burning basically stops for \texttt{WD}s. The evolution of \texttt{WD}s is a cooling process, together with a contraction process early on. Those \texttt{WD}s with a hydrogen(H)-rich atmosphere are classified as DA-tpye \texttt{WD}s. Those \texttt{WDs} with a helium(He)-rich atmosphere and an ionized He-rich atmosphere are classified as DB-tpye \texttt{WD}s and DO-type \texttt{WD}s respectively. DA-type \texttt{WDs} account for roughly 80\% of the total number of \texttt{WD}s (Bischoff-Kim \& Metcalfe 2011).

There are 260 ZZ Ceti stars (also named DA-type variable stars, DAV stars for short) observed and identified (C$\acute{o}$rsico 2020). This provides rich target sources for the study of DAV stars. The pulsation instability strip for DAV stars is basically from 12270\,K to 10850\,K (Gianninas, Bergeron \& Fontaine 2005, Gianninas, Bergeron \& Ruiz 2011). The driving mechanism of pulsating modes for DAV stars are the classical $\kappa$-$\gamma$ driving mechanism (Dolez \& Vauclair 1981, Winget et al. 1982) and the convective driving mechanism (Brickhill 1991, Goldreich \& Wu 1999). The pulsation modes are very sensitive to the stellar structure of DAV stars and therefore can be used to study the internal structure of DAV stars.

L19-2 was found to be the 11th DAV star by McGraw in 1977 (McGraw 1977) at the Sutherland observing station of the South African Astronomical Observatory (SAAO). It is a very southerly star with a declination of -81 degrees. McGraw (1977) reported two low amplitude pulsation modes near 192\,s and 113\,s for L19-2. Based on 292 hours of photometric observation at the Sutherland observing station of the SAAO, O'Donoghue $\&$ Warner (1982) obtained five independent modes of 350\,s, 192\,s, 143\,s, 118\,s, and 113\,s for L19-2. These five modes have frequency splitting phenomena, which is helpful to carry out reliable mode identifications. O'Donoghue $\&$ Warner (1987) observed L19-2 intermittently from 1983 July to 1985 February for 26 hours and studied the rates of period change for the modes of 192\,s and 113\,s. L19-2 was observed in a Whole Earth Telescope (Nather et al. 1990) campaign, XCOV12 in 1995, and in a single site from 1994 to 1997 (Sullivan 1998). L19-2 was observed in Mt John University Observatory (MJUO) in New Zealand from 1994 to 2013 (Sullivan \& Chote 2015). Based on the observed data from 1976 to 2013 for L19-2, Sullivan \& Chote (2015) derived the rates of period change of (3.0 $\pm$ 0.6) $\times$ $10^{-15}$ s/s for the modes of 192\,s and 113\,s through the O-C method.

L19-2 has a long observation history and has been widely studied in asteroseismology, such as Bradley (2001), Castanheira \& Kepler (2009), and Romero et al. (2012). Bischoff-Kim \& Montgomery (2018) reported a latest version of the White Dwarf Evolution Code (\texttt{WDEC}), which realized the parameterization of the oxygen(O) profile in the central core of a \texttt{WD}. The calculated pulsation modes are very sensitive to the element abundances in the central core. The \texttt{WDEC} (2018 version) has an opportunity to achieve more accurate fitting results of asteroseismological models. We perform an asteroseismological study on the DAV star L19-2 based on the \texttt{WDEC} (2018 version). In Sect. 2, input physics and model calculations are displayed. In Sect. 3, we show the asteroseismology of L19-2, including the model fittings, the comparison to previous spectral and asteroseismological results, and the rate of period change. At last, we give a discussion and conclusions in Sect. 4.


\section{Input Physics and Model Calculations}

\texttt{WDEC} evolves white dwarf models from $\sim$\,100\,000\,K to a specific effective temperature ($T_{\rm eff}$) we require, which is fast and versatile to evolve grids of \texttt{WD} models. The \texttt{WDEC} (2018 version) can evolve C/O core \texttt{WD}s and He core \texttt{WD}s. When evolving the C/O core \texttt{WD}s, we can run the quick and dirty old version, the state of the art version v15, and the state of the art version v16. The quick and dirty old version is the previous version of \texttt{WDEC} (Kutter \& Savedoff 1969, Lamb \& van Horn 1975, Wood 1990). The state of the art version uses the equation of state tables and opacities tables from the Modules for Experiments in Stellar Astrophysics (\texttt{MESA}, Paxton et al. 2011 and their follow up papers, version r8118). The v15 version allows a direct input of an O profile. For the v16 version, the O profile is parameterized by six parameters. We use the state of the art version v16 to evolve grids of DAV star models. The standard mixing length theory (B$\ddot{o}$hm \& Cassinelli 1971) is used and the mixing length parameter is adopted as 0.6 (Bergeron et al. 1995).

We show the parameter spaces of evolved \texttt{WD}s in Table 1. $M_{*}$ is the stellar mass, $M_{\rm env}$ is the envelope mass, $M_{\rm He}$ is the He layer mass, and $M_{\rm H}$ is the H atmosphere mass. $X_{\rm He}$ is the helium abundance in the mixed C/He/H region. The parameters of h1-3 and w1-3 refer to the O profile in the core. The parameter h1 is the O abundance in the core center, while w1 is the mass fraction of $X_{O}$ = h1. The parameter h2 and h3 correspond to the O abundance of two knee points on the reducing O profile. The w2 and w3 correspond to the masses of the gradient regions of O profile. For more details about the core parameters, see Fig. 1 of Bischoff-Kim (2018) and the user manual of the procedure (Bischoff-Kim \& Montgomery 2018).

The \texttt{WDEC} (2018 version) takes about 12 seconds to evolve a \texttt{WD}. For the initial ranges and crude steps in Table 1, 1\,009\,908 \texttt{WD} models need to be evolved. It took us almost half a year to finish the model evolutions. With the preliminary fitting results, we will use the fine steps and sophisticated steps to optimize the fitting models.

\begin{table*}
\begin{center}
\caption{The explored parameter spaces of \texttt{WD}s evolved by \texttt{WDEC}.}
\begin{tabular}{llllllllll}
\hline
Parameters                             &Initial ranges     &Crude steps    &Fine steps     &Sophisticated steps  &Optimal values          \\
\hline
$M_{*}$/$M_{\odot}$                    &[0.500,0.850]      &0.010          &0.005          &0.005                &0.800$\pm$0.003         \\
$T_{\rm eff}$(K)                       &[10600,12600]      &250            &50             &10                   &12360$\pm$18            \\
-log($M_{\rm env}/M_{\rm *}$)          &[1.500,3.000]      &1.000          &0.003          &0.001                &2.405$\pm$0.074         \\
-log($M_{\rm He}/M_{\rm *}$)           &[3.000,5.000]      &1.000          &0.003          &0.001                &2.972$\pm$0.019         \\
-log($M_{\rm H}/M_{\rm *}$)            &[5.000,10.000]     &1.000          &0.003          &0.001                &5.030$\pm$0.004         \\
$X_{\rm He}$ in mixed C/He/H region    &[0.100,0.900]      &0.160          &0.030          &0.001                &0.128$\pm$0.015         \\
\hline
$X_\mathrm{O}$ in the core             &                   &               &               &                     &                        \\
\hline
h1                                     &[0.600,0.750]      &0.030          &0.010          &0.001                &0.914$\pm$0.012         \\
h2                                     &[0.650,0.710]      &0.030          &0.010          &0.001                &0.672$\pm$0.004         \\
h3                                     &0.850              &               &0.010          &0.001                &0.848$\pm$0.012         \\
w1                                     &0.350              &               &0.010          &0.001                &0.413$\pm$0.001         \\
w2                                     &0.480              &               &0.010          &0.001                &0.427$\pm$0.001         \\
w3                                     &0.090              &               &0.010          &0.001                &0.140$\pm$0.002         \\
\hline
\end{tabular}
\end{center}
\end{table*}

\section{Asteroseismology of L19-2}

O'donoghue \& Warner (1982) reported frequency splitting phenomena among modes of 350\,s, 192\,s, 143\,s, 118\,s, and 113\,s, as shown in Table 2. They were also detected by the 1995 WET run (Sullivan \& Chote 2015). For the frequency splitting values of Sullivan \& Chote (2015), (13.0 + 12.9 + 11.1)\,$\mu$Hz divided by 3 is 12.3\,$\mu$Hz and (18.9 + 39.6)\,$\mu$Hz divided by 3 is 19.5\,$\mu$Hz. The ratio between 12.3 and 19.5 is 0.63, which is close to theoretical value of $\frac{\delta \nu_{l=1}}{\delta \nu_{l=2}}$ = 0.6 (Brickhill 1975, Winget et al. 1991). Therefore, we adopt the central modes and the high-amplitude modes of 350.15\,s, 192.61\, 143.42\,s, 118.68\,s, and 113.78\,s to constrain the fitting models. The modes of 350.15\,s, 192.61\, and 118.68\,s are identified as $l$ = 1 and $m$ = 0 modes, while the modes of 143.42\,s and 113.78\,s are identified as $l$ = 2 and $m$ = 0 modes. The five modes were also fitted by Castanheira \& Kepler (2009) and Romero et al. (2012). In addition, $\delta \nu_{l=1}$ = 12.3\,$\mu$Hz for $l$ = 1 modes corresponds to a rotational period around 11.29 hours.

\begin{table*}
\begin{center}
\caption{The observed modes for L19-2 by O'donoghue \& Warner (1982) and Sullivan \& Chote (2015). The mark '?' denotes that the existence of the component is not well established by O'donoghue \& Warner (1982).}
\begin{tabular}{llllllllllll}
\hline
\multicolumn{4}{c}{O'donoghue \& Warner (1982)}&\multicolumn{4}{c}{Sullivan \& Chote (2015)}                                      &Selected                  \\

Freq.        &Ampl.       &$\delta$Freq. &Peri.                &Freq.               &Ampl.       &$\delta$Freq. &Peri.            &$P_{\rm obs}$ ($l$,$m$)   \\
\hline
($\mu$Hz)    &(mma)       &($\mu$Hz)     &(s)                  &($\mu$Hz)           &(mma)       &($\mu$Hz)     &(s)              &(s)                       \\
\hline
2855.932$\pm$0.010  &1.00        &              &350.1484$\pm$0.0012  &2855.9       &0.77        &              &350.15           &350.15 ($l$=1, $m$=0)     \\
                    &            &11.626        &                     &             &            &              &                 &                          \\
2867.558            &0.43        &              &348.7288             &             &            &              &                 &                          \\
5178.857            &0.81        &              &193.0928             &5178.8       &0.95        &              &193.09           &                          \\
                    &            &12.992        &                     &             &            &13.0          &                 &                          \\
5191.849$\pm$0.010  &6.00        &              &192.6096$\pm$0.0004  &5191.8       &5.92        &              &192.61           &192.61 ($l$=1, $m$=0)     \\
                    &            &12.993        &                     &             &            &12.9          &                 &                          \\
5204.842            &0.73        &              &192.1288             &5204.7       &0.61        &              &192.13           &                          \\
6972.571$\pm$0.010  &0.55        &              &143.4191$\pm$0.0002  &6972.5       &0.49        &              &143.42           &143.42 ($l$=2, $m$=0)     \\
                    &            &18.534        &                     &             &            &18.9          &                 &                          \\
6991.105?           &0.26        &              &143.0389             &6991.4       &0.41        &              &143.03           &                          \\
8408.116?           &0.24        &              &118.9327             &             &            &              &                 &                          \\
                    &            &18.200        &                     &             &            &              &                 &                          \\
8426.316            &1.10        &              &118.6758             &8426.3       &1.34        &              &118.68           &118.68 ($l$=1, $m$=0)     \\
                    &            &11.143        &                     &             &            &11.1          &                 &                          \\
8437.459$\pm$0.010  &1.81        &              &118.5191$\pm$0.0001  &8437.4       &1.72        &              &118.52           &                          \\
8757.633?           &0.27        &              &114.1861             &             &            &              &                 &                          \\
                    &            &31.420        &                     &             &            &              &                 &                          \\
8789.053$\pm$0.010  &2.21        &              &113.7779$\pm$0.0001  &8789.1       &2.05        &              &113.78           &113.78 ($l$=2, $m$=0)     \\
                    &            &39.640        &                     &             &            &39.6          &                 &                          \\
8828.693            &0.52        &              &113.26705            &8828.7       &0.43        &              &113.27           &                          \\
\hline
\end{tabular}
\end{center}
\end{table*}

\subsection{The Model Fittings}

During the initial fittings, an average fitting error is generally greater than 1 second. We use a root-mean-square residual $\sigma_{RMS}$ to evaluate the fitting results.
\begin{equation}
\sigma_{RMS}=\sqrt{\frac{1}{n} \sum_{n}(P_{\rm obs}-P_{\rm cal})^2}.
\end{equation}
\noindent In Eq.\,(1), $n$ is the number of fitted observed modes, which is 5 for L19-2. Then we gradually reduce the parameter steps around the models with minimum $\sigma_{RMS}$ values. The ranges of each parameters are reduced. For the fine steps, there are 5 grid points for each parameters. For the sophisticated steps, there are 3 grid points for each parameters. There are 6 global parameters and 6 $X_\mathrm{O}$ parameters in Table 1. We fixed the $X_\mathrm{O}$ parameters and make a grid of global parameters to reduce the fitting error. Then fix the global parameters and make a grid of $X_\mathrm{O}$ parameters to reduce the fitting error. By repeating these two processes over and over again, the fitting model we choose will become closer and closer to an optimal fitting model. It took us almost a month to find the best fitting model for L19-2. When the $\sigma_{RMS}$ is less than 1 second, we use an absolute difference $\Phi$ to select the optimal model.
\begin{equation}
\Phi=\frac{1}{n} \sum_{n}|P_{\rm obs}-P_{\rm cal}|
\end{equation}
\noindent In Eq.\,(2), $n$ is 5 for L19-2, the same as in Eq.\,(1). At last, we obtain an optimal model with $\Phi$ = 0.06\,s. The parameters of the optimal model are displayed in the last column of Table 1. We use a half height and full width of the reciprocal of n$\Phi$ to calculate each parameter error. Note the sophisticated steps in the penultimate column, the optimal model comes from a very fine fitting work. It is a massive and hot DAV star with relatively a thick H atmosphere and a thick He layer.

In Fig. 1, we show a color figure $\Phi$ to stellar mass and effective temperature. The colors indicate the fitting error $\Phi$ in Eq.\,(2). The abscissa shows the stellar mass from 0.500\,$M_{\odot}$ to 0.850\,$M_{\odot}$ with a step of 0.005\,$M_{\odot}$. The ordinate shows the effective temperature from 10\,600\,K to 12\,600\,K with a step of 10\,K. The other parameters are fixed to the parameters of the optimal model, the last column in Table 1. There are 14\,271 DAV star models in Fig. 1. We can see that the optimal model has a massive stellar mass and a hot effective temperature. In Fig. 2, we show a color figure $\Phi$ to H atmosphere mass and envelope mass. There are 13\,311 DAV star models in Fig. 2. We can see that the optimal model has a thick H atmosphere mass. Compared with other parameters, the sensitivity of fitting error to the envelope mass is relatively low. It is consistent with the large error of the envelope mass in the last column of Table 1. The optimal model has -log($M_{\rm env}/M_{\rm *}$) of 2.405$\pm$0.074. In Fig. 3, we show a sensitivity figure of 8 parameters of the optimal model. The abscissa is the value of 8 parameters respectively, and the ordinate is the fitting error $\Phi$. When studying the sensitivity of each parameter, other parameters are fixed at the values of the optimal model. In fact, for the optimal model, every parameter is optimal.

\begin{figure}
\begin{center}
\includegraphics[width=9.0cm,angle=0]{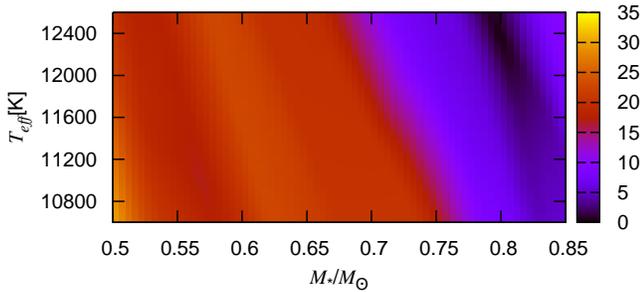}
\end{center}
\caption{Color figure of $\Phi$ to stellar mass and effective temperature. There are 14\,271 DAV star models.}
\end{figure}

\begin{figure}
\begin{center}
\includegraphics[width=9.0cm,angle=0]{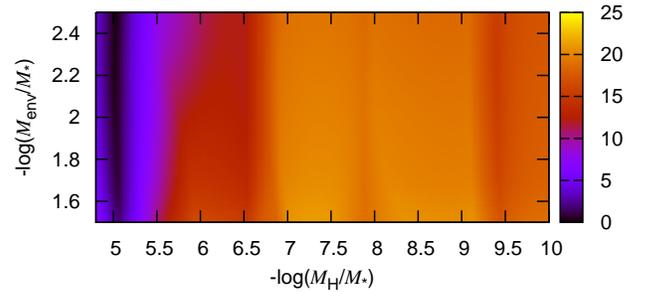}
\end{center}
\caption{Color figure of $\Phi$ to H atmosphere mass and envelope mass. There are 13\,311 DAV star models.}
\end{figure}

\begin{figure*}
\begin{center}
\includegraphics[width=18.0cm,angle=0]{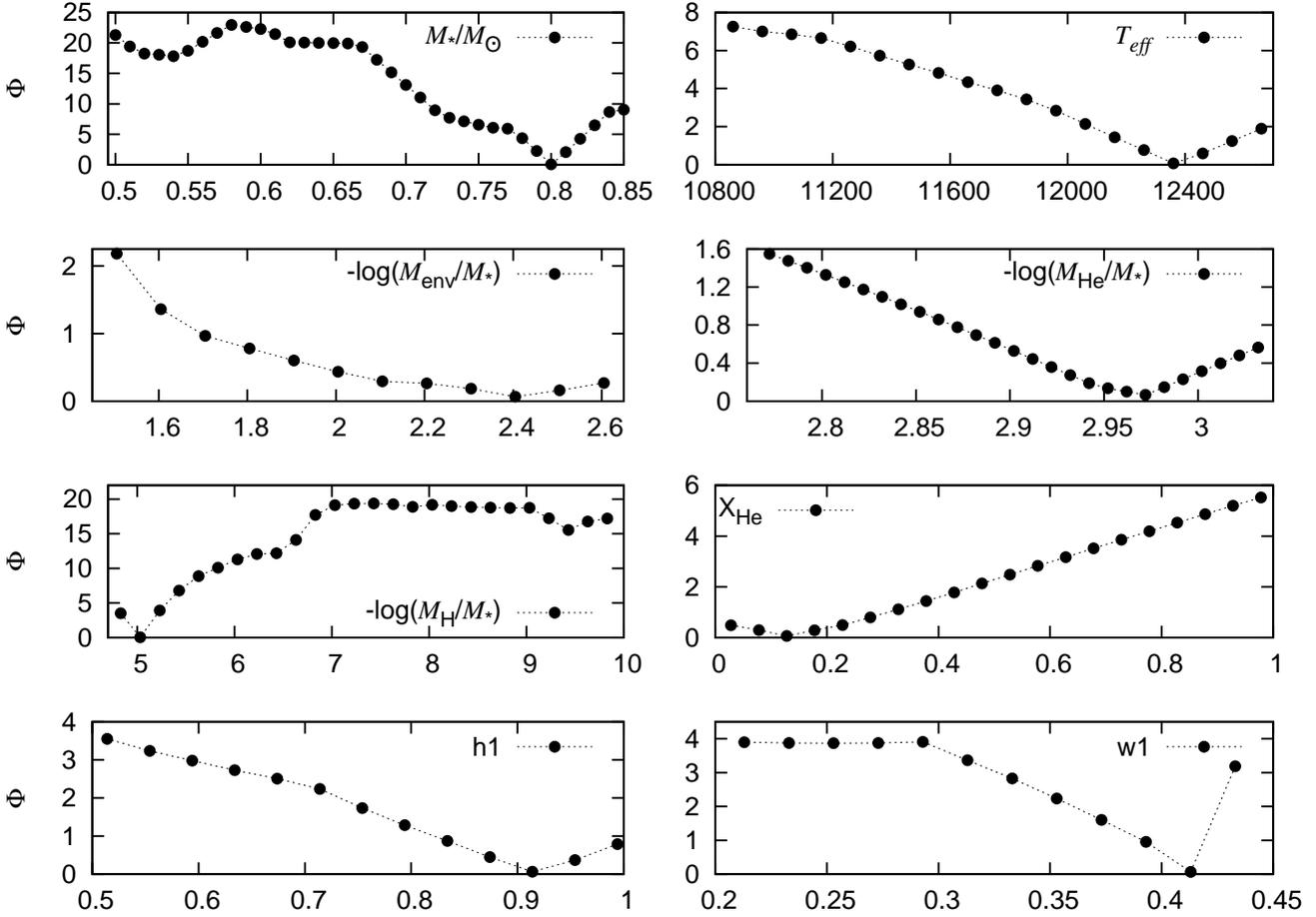}
\end{center}
\caption{Sensitivity figure of 8 parameters of the optimal model. The abscissa is the value of 8 parameters respectively, and the ordinate is the fitting error $\Phi$.}
\end{figure*}

\begin{figure}
\begin{center}
\includegraphics[width=9.0cm,angle=0]{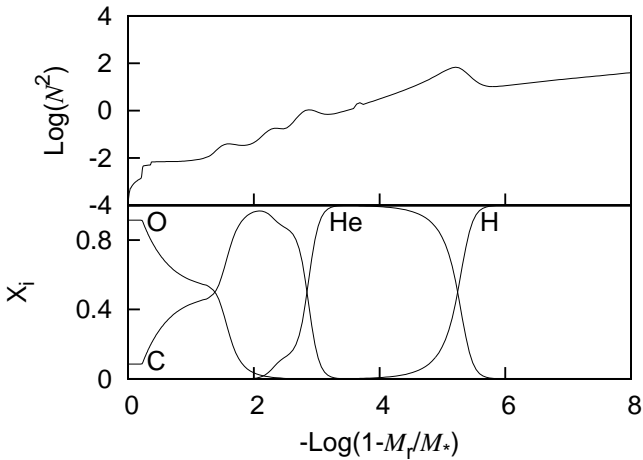}
\end{center}
\caption{The core composition profiles and corresponding Brunt-V\"ais\"al\"a frequency for the optimal model.}
\end{figure}

In Fig. 4, we show the core composition profiles and corresponding Brunt-V\"ais\"al\"a frequency for the optimal model. The composition gradient zones in the low panel result in bumps in the up panel. There is a thin convection zone on the hydrogen atmosphere surface. In Table 1 and Fig. 4, we can see that the central O abundance h1 is 0.914. It seems too large. The calculation of stellar evolution shows that the central O abundance of a \texttt{WD} is about 0.5-0.8 (Salaris et al. 2010, Althaus et al. 2010). The central O abundance of a \texttt{WD} is affected by the progenitor main sequence core overshooting, the initial metal abundance, the semiconvection during the central He burning, the breathing pulse suppression, and the specific C/O nuclear reaction rate. The asteroseismological research of \texttt{WD}s has an opportunity to explore these physical properties of progenitor stars. Bischoff-Kim (2018) fitted a DAV star R548 and obtained a best fitting model with central O abundance h1 of 0.90. Kalup Bogn$\acute{a}$r \& S$\acute{o}$dor (2021) fitted a DAV star HS 1625+1231 and obtained a best fitting model with central O abundance h1 of 0.5 or 0.9. Giammichele et al. (2018) fitted a DBV star KIC08626021 and obtained a best fitting model with central O abundance of 0.86. Duan et al. (2021) fitted a DBV star EPIC 228782059 and obtained a best fitting model with central O abundance h1 of 0.52. More asteroseismological work on \texttt{WD}s is required to study the central O abundance of \texttt{WD}s.

\begin{table*}
\begin{center}
\caption{The calculated periods and the fitting result of the optimal model. $P_{\rm cal}(l,k)$ is the calculated modes for the optimal model. $P_{\rm obs}$ is the observed modes for L19-2.}
\begin{tabular}{lllllllllllll}
\hline
$P_{\rm cal}(l,k)$&$P_{\rm obs}$&$P_{\rm cal}(l,k)$&$P_{\rm obs}$&$P_{\rm cal}(l,k)$&$P_{\rm obs}$&$P_{\rm cal}(l,k)$&$P_{\rm obs}$&$P_{\rm cal}(l,k)$&$P_{\rm obs}$\\
(s)               &(s)          &(s)               &(s)          &(s)               &(s)          &(s)               &(s)          &(s)               &(s)\\
\hline
118.68(1,1)       &118.68       &766.79(1,19)      &             &114.10(2,2)       &113.78       &470.26(2,20)      &             &832.48(2,38)      &   \\
192.61(1,2)       &192.61       &806.98(1,20)      &             &137.70(2,3)       &             &493.31(2,21)      &             &854.14(2,39)      &   \\
212.03(1,3)       &             &828.71(1,21)      &             &143.42(2,4)       &143.42       &509.86(2,22)      &             &874.18(2,40)      &   \\
242.77(1,4)       &             &862.66(1,22)      &             &163.66(2,5)       &             &528.00(2,23)      &             &897.19(2,41)      &   \\
281.22(1,5)       &             &903.80(1,23)      &             &192.47(2,6)       &             &552.82(2,24)      &             &916.92(2,42)      &   \\
324.65(1,6)       &             &929.70(1,24)      &             &205.58(2,7)       &             &571.85(2,25)      &             &938.17(2,43)      &   \\
350.15(1,7)       &350.15       &963.37(1,25)      &             &228.49(2,8)       &             &586.97(2,26)      &             &957.71(2,44)      &   \\
390.68(1,8)       &             &1006.79(1,26)     &             &254.21(2,9)       &             &608.41(2,27)      &             &977.29(2,45)      &   \\
420.36(1,9)       &             &1042.87(1,27)     &             &266.83(2,10)      &             &627.94(2,28)      &             &999.04(2,46)      &   \\
450.86(1,10)      &             &1076.11(1,28)     &             &289.71(2,11)      &             &648.21(2,29)      &             &1016.83(2,47)     &   \\
493.19(1,11)      &             &1112.59(1,29)     &             &313.06(2,12)      &             &669.77(2,30)      &             &1039.99(2,48)     &   \\
519.57(1,12)      &             &1147.33(1,30)     &             &329.96(2,13)      &             &692.30(2,31)      &             &1060.30(2,49)     &   \\
552.97(1,13)      &             &1178.14(1,31)     &             &352.76(2,14)      &             &711.39(2,32)      &             &1081.99(2,50)     &   \\
583.71(1,14)      &             &                  &             &366.51(2,15)      &             &732.24(2,33)      &             &1104.20(2,51)     &   \\
620.46(1,15)      &             &                  &             &386.65(2,16)      &             &754.68(2,34)      &             &1124.21(2,52)     &   \\
663.44(1,16)      &             &                  &             &412.42(2,17)      &             &770.14(2,35)      &             &1146.90(2,53)     &   \\
689.49(1,17)      &             &                  &             &424.91(2,18)      &             &791.55(2,36)      &             &1164.78(2,54)     &   \\
723.07(1,18)      &             &68.53(2,1)        &             &444.70(2,19)      &             &811.55(2,37)      &             &1185.96(2,55)     &   \\
\hline
\end{tabular}
\end{center}
\end{table*}

In Table 3, we show the calculated periods and the detailed fitting result of the optimal model. The modes of 118.68\,s, 192.61\,s, 350.15\,s, and 143.42\,s are accurately fitted. The mode of 113.78\,s is fitted by 114.10\,s, with an error of 0.32\,s, and therefore $\Phi$ is 0.06\,s for the optimal model. By slightly adjusting the parameters in Table 1 repeatedly, an optimal fitting model with the fitting error of $\Phi$ = 0.06\,s can be obtained.

The radial order $k$ of the modes of 113.78\,s and 143.42\,s was identified as 2 and 4 respectively by Bradley (2001), and 2 and 3 respectively by Romero et al. (2012). For our optimal model, that is 2 and 4 respectively. For the work of Bradley (2001) and Romero et al. (2012), the radial order $k$ of modes of 118.68\,s, 192.61\,s, and 350.15\,s is identified as 1, 2, and 6 respectively. For our optimal model, that is 1, 2, and 7 respectively. Actually, the asymptotic theory for $g$-modes is suitable for high $k$ modes (Tassoul 1980). It is very common that the extreme low $k$ modes do not satisfy the asymptotic theory for $g$-modes (Brassard et al. 1992).

\subsection{The Comparison to Previous Spectral and Asteroseismological Results}

For our optimal model, the luminosity is Log(L/$L_{\odot}$) = -2.676. The bolometric magnitude of the Sun $M_{bol,\odot}$ is adopted as 4.74 (Cox 2000). Using the correction $M_{bol}$ = $M_{bol,\odot}$ - 2.5$\times$Log(L/$L_{\odot}$), we can derive the bolometric magnitude of the optimal model $M_{bol}$ = 11.43. In Table 1 of Bergeron, Wesemael \& Beauchamp (1995), the calibration of BC(V) is -0.441, -0.611, and -0.828 for the DA-type star model with $T_{\rm eff}$ = 11000\,K, 12000\,K, and 13000\,K respectively. For our optimal model with $T_{\rm eff}$ = 12360\,K, we estimate a BC(V) of -0.689. Then the absolute visual magnitude of the optimal model is $M_{V}$ = $M_{bol}$ - BC(V) = 12.119. The visual magnitude of L19-2 is $m_{V}$ = 13.75 (Bergeron et al. 1995). Applying the formula 5log$d$ = $m_{V}$ + 5 - $M_{V}$, we can derive a distance for L19-2 with $d$ = 21.18\,pc. The distance is corresponding to a triangular parallax of 47.21 mas. In Global Astrometric Interferometer for Astrophysics (Gaia) Data Release 2 (DR2), the parallax of L19-2 is 47.7874 $\pm$ 0.0295 (Gaia collaboration et al. 2018). In other words, for L19-2, the distance measured according to our optimal model has only 1\% error compared with the observed distance from Gaia DR2. For our optimal model, the period fitting to the observed modes and the distance matching to the Gaia DR2 data are self consistent.

\begin{table*}
\begin{center}
\caption{Diagram of best-fitting models for L19-2. The ID number 1, 2, 3, 4, 5, 6, and 7 is the spectral work of Bergeron et al. (2004), Koester \& Allard (2000), the asteroseismological work of Bradley (2001), Castanheira \& Kepler (2009), Romero et al. (2012), Romero et al. (2013), and the present work respectively.}
\begin{tabular}{lllllccccc}
\hline
ID         &$T_{\rm eff}$&log\,$g$     &$M_{*}$         &log($M_{\rm H}/M_{*}$)   &log($M_{\rm He}/M_{*}$)&$\Phi$         \\
           &(K)          &             &($M_{\odot}$)   &                         &                       &(s)            \\
\hline
1          &12100$\pm$200&8.21$\pm$0.05&                &                         &                       &               \\
2          &12150$\pm$100&8.15$\pm$0.15&                &                         &                       &               \\
3          &12300$\pm$250&8.19         &0.72            &-4.0                     &-2.05                  &1.3            \\
4          &12100        &             &0.75            &-4.5                     &-2.0                   &3.8            \\
5          &12105$\pm$360&8.17$\pm$0.07&0.705$\pm$0.033 &-4.44$^{+0.16}_{-0.27}$  &-2.12                  &1.224          \\
6          &12033$\pm$316&8.17$\pm$0.08&0.705$\pm$0.023 &-4.44                    &-2.12                  &1.220          \\
7          &12360$\pm$18 &8.332        &0.800$\pm$0.003 &-5.030$\pm$0.004         &-2.972$\pm$0.019       &0.06           \\
\hline
\end{tabular}
\end{center}
\end{table*}

In Table 4, we list the best-fitting models of previous spectral and asteroseismological work together with the present work. The ID number 1 and 2 are from the spectral work of Bergeron et al. (2004) and  Koester \& Allard (2000) respectively. Based on the previous old \texttt{WDEC} procedure, Bradley (2001) evolved grids of DAV star models with a 20:80 C/O core and did asteroseismological studies on L19-2 and GD165. The best-fitting model for L19-2 is the ID 3. Based on the previous old \texttt{WDEC} procedure, Castanheira \& Kepler (2009) evolved grids of DAV star models with a homogeneous 50:50 C/O core and did asteroseismological studies on 83 DAV stars. They obtained an average H atmosphere with log($M_{\rm H}/M_{*}$) = -6.3 for DAV stars. Their best-fitting model for L19-2 is the ID 4. With the \texttt{LPCODE} evolutionary code (Althaus et al. 2005, 2010, Renedo et al. 2010), Romero et al. (2012) evolved fully evolutionary DAV star models and did asteroseismological studies on 44 bright DAV stars. Their best-fitting model for L19-2 is the ID 5. With the \texttt{LPCODE}, Romero et al. (2013) evolved fully evolutionary DAV star models and did asteroseismological studies on 42 massive DAV stars. Their fully evolutionary DAV star models considered the physical process of crystallization. The best-fitting model for L19-2 is the ID 6. ID 7 is the optimal model of the present work.

All the best-fitting models show that L19-2 is a hot DAV star. It has a relatively large stellar mass corresponding to a relatively large gravitational acceleration. It has relatively a thick H atmosphere and a thick He layer. Our optimal model has the smallest value of $\Phi$. By parameterizing the core oxygen profile, the \texttt{WDEC} (2018 version) procedure can greatly reduce the fitting error of asteroseismological models. It seems that the other parameters of our optimal model are a little modifications to that of the previous work. In addition, the best-fitting model of ID 6 taking the crystallization effect into account is very similar to that of ID 5.

\subsection{The Rate of Period Change}

In this subsection, we discuss the trapped modes and the rate of period change for L19-2. Winget, van Horn, \& Hansen (1981) first proposed that the composition transition zone could be used as a boundary condition to form standing waves. Therefore, some modes are trapped in H atmosphere and some modes are trapped in He layer. Trapped modes have small amplitude in the core, and therefore have small kinetic energy of oscillation, $E_{kin}$, as described by Winget, van Horn, \& Hansen (1981). In Fig. 5, we show the kinetic energy of the calculated modes for the optimal model. We can see that the $l$ = 1 and $k$ = 1 mode is obviously a trapped mode. Namely, the observed mode of 118.68\,s is a trapped mode. It is consistent with the discussion performed by C$\acute{o}$rsico et al. (2016).

\begin{table*}
\caption{Comparing the calculated values of $\dot{P}$ of the optimal model to the observed values of that obtained through O-C method for L19-2.}
\begin{center}
\begin{tabular}{llllllll}
\hline
\multicolumn{2}{c}{Observed values}&\multicolumn{2}{c}{Calculated from the optimal model}&\multicolumn{2}{c}{Calculated by C$\acute{o}$rsico et al. (2016)}\\
\hline
$P_{\rm obs}$ &$\dot{P}_{\rm obs}$ & $P_{\rm cal}$($l$,$k$) &$\dot{P}_{\rm cal}$($T_{\rm eff}$ interval of 200K, 400K, 600K)& $P_{\rm cal}$($l$,$k$) &$\dot{P}_{\rm cal}$           \\
\hline
[s]           & [$10^{-15}$s/s]    & [s]                    &[$10^{-15}$s/s]                     & [s]                    &[$10^{-15}$s/s]               \\
\hline
118.68        &                    &118.68(1,1)             &0.58$\pm$0.00\,(0.58,\,0.58,\,0.58) &117.21(1,1)             &0.47                          \\
192.61        &3.0$\pm$0.6         &192.61(1,2)             &2.72$\pm$0.02\,(2.74,\,2.70,\,2.74) &192.88(1,2)             &2.41                          \\
350.15        &                    &350.15(1,7)             &2.80$\pm$0.01\,(2.81,\,2.79,\,2.81) &                        &                              \\
113.78        &3.0$\pm$0.6         &114.10(2,2)             &1.46$\pm$0.01\,(1.46,\,1.46,\,1.47) &113.41(2,2)             &1.42                          \\
143.42        &                    &143.42(2,4)             &1.93$\pm$0.03\,(1.96,\,1.90,\,1.95) &145.41(2,3)             &0.89                          \\
\hline
\end{tabular}
\end{center}
\end{table*}

\begin{figure}
\begin{center}
\includegraphics[width=9.0cm,angle=0]{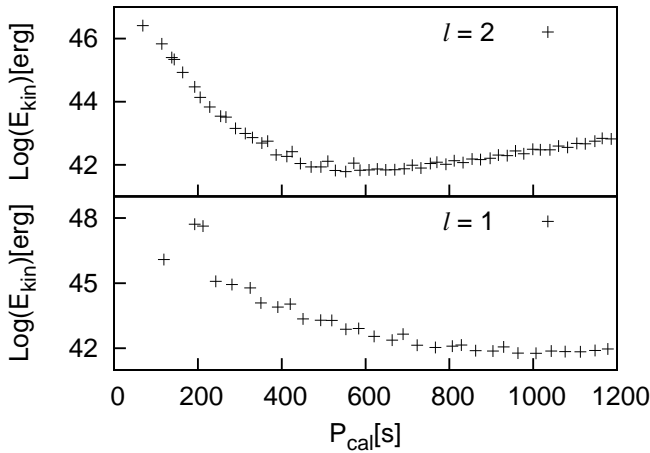}
\end{center}
\caption{The kinetic energy of the calculated modes for the optimal model. The radial order $k$ is from 1 to 31 for $l$ = 1 modes and from 1 to 55 for $l$ = 2 modes.}
\end{figure}

The theoretical calculation shows that the rate of period change for a trapped mode is smaller than that of other modes (Bradley 1996). This is because that the oscillation energy of the trapped mode is concentrated on the surface of a \texttt{WD}, and the residual gravitational contraction effect on the surface can not be completely ignored. For our optimal model, we calculate the rate of period change for the calculated modes through the following equation.
\begin{equation}
\dot{P}(k)=\frac{P_{2}(k)-P_{1}(k)}{Age_{2}-Age_{1}}.
\end{equation}
\noindent In Eq.\,(3), $P_{2}(k)$ and $P_{1}(k)$ are the pulsation periods of the same mode (same $l$ and same $k$) at two near epochs. The two epochs include $T_{\rm eff}$ from 12460\,K to 12260\,K, from 12560\,K to 12160\,K, and from 12660\,K to 12060\,K respectively. The other parameters are fixed to the parameters of the optimal model. The $T_{\rm eff}$ has an interval of 200\,K, 400\,K, and 600\,K respectively. According to Eq.\,(3), we derive three rates of period change for the calculated modes and obtain an average value for each mode, which are listed in Table 5 (column 4). The mode of 118.68\,s is a trapped one and then has a small rate of period change.

The observed minus calculated times of maxima (O-C) method is widely used to measure the rate of period change of very stable pulsation signals, once the signals have been observed during a long time interval. The O-C method is derived from the Taylor's theorem in mathematics (Costa \& Kepler 2008). Based on the observation data from 1974 to 2010 for a DAV star G117-B15A, using the O-C method, Kepler (2012) report that the rate of period change for a stable mode of 215\,s is (4.19 $\pm$ 0.73) $\times$ $10^{-15}$ s/s, taking the proper motion correction into account. Based on the observation data from 1970 to 2012 for a DAV star R548, using the O-C method, Mukadam et al. (2013) report that the rate of period change for a stable mode of 213\,s is (3.3 $\pm$ 1.1) $\times$ $10^{-15}$ s/s, taking the proper motion correction into account. Based on the observation data from 1976 to 2013 for L19-2, using the O-C method, Sullivan \& Chote (2015) report that the rates of period change for both the stable modes of 192\,s and 113\,s are (4.0 $\pm$ 0.6) $\times$ $10^{-15}$ s/s. The proper motion correction for L19-2 is about -1.0 $\times$ $10^{-15}$ s/s (Pajdosz 1995). The observed rates of period change for both modes of 192\,s and 113\,s are (3.0 $\pm$ 0.6) $\times$ $10^{-15}$ s/s, which are listed in Table 5.

In Table 5, the calculated rates of period change for both the modes of 192\,s and 113\,s are smaller than the observed ones. Taking the uncertainties into account, the $\dot{P}_{\rm cal}$ for the mode of 192\,s is basically consistent with the $\dot{P}_{\rm obs}$. However, the $\dot{P}_{\rm cal}$ for the mode of 113\,s is still smaller than the $\dot{P}_{\rm obs}$. With the best-fitting model of ID 6 in Table 4, C$\acute{o}$rsico et al. (2016) studied the observed and calculated periods and the rates of period change. For the mode of 192\,s and 113\,s, C$\acute{o}$rsico et al. (2016) calculated the theoretical rate of period change as 2.41 $\times$ $10^{-15}$ s/s and 1.42 $\times$ $10^{-15}$ s/s respectively, as shown in the last column of Table 5. They discussed the sources of uncertainties in theoretical rates of period change, including the $^{12}C(\alpha,\gamma)^{16}O$ reaction rate, the asteroseismological model and the period-fit procedure, the element mixing during stellar evolution, the progenitor metallicity and rotation, and the possible axion emission (C$\acute{o}$rsico et al. 2016). The extra sources of cooling such as axions lead to a higher rates of period change. Therefore, the smaller part that the theoretical value is smaller than the observed value has an opportunity to limit the mass of the axion.

For the modes of 192\,s, and 113\,s, the theoretical rates of period change (column 4) in Table 5 are slightly larger but basically consistent with that of C$\acute{o}$rsico et al. (2016) (column 6 in Table 5) when fitting the observed values. The mode of 118\,s is a trapped one and has an obviously small $\dot{P}_{\rm cal}$ for our optimal model and that of C$\acute{o}$rsico et al. (2016). C$\acute{o}$rsico et al. (2016) fit the 143\,s mode with a partially trapped mode of 145.41\,s ($l$ = 2 and $k$ = 3) and obtain a smaller theoretical $\dot{P}_{\rm cal}$. Our optimal model with the smallest $\Phi$ has an opportunity to reduce the uncertainty from the asteroseismological model and the period-fit procedure when calculating the theoretical $\dot{P}_{\rm cal}$.

\section{Discussion and Conclusions}

In this paper, we first summarize the photometric observations and mode identifications of the DAV star L19-2. L19-2 was intermittently observed from 1976 to 2013, and five independent modes were identified. The five modes are 3 $l$ = 1 modes (118\,s, 192\,s, and 350\,s) and 2 $l$ = 2 modes (113\,s and 143\,s). Then, we evolve 1\,009\,908 DAV star models and calculate the theoretical pulsation periods by \texttt{WDEC} (2018 version) to fit the 5 observed modes of L19-2. It is a large sample work, which took about half a year. We gradually reduce the parameter steps and the parameter ranges according to the fitting results, and repeatedly evolve DAV star models to find an optimal one. The sophisticated steps are used to select the optimal model, as shown in Table 1.

The central O abundance of the optimal model in Table 1 is 0.914, which is significantly higher than the results of the present stellar structure and evolution theory (around 0.5-0.8 from Salaris et al. 2010 and Althaus et al. 2010), although some recent works have also obtained high central O abundance (Bischoff-Kim 2018, Kalup Bogn$\acute{a}$r \& S$\acute{o}$dor 2021, Giammichele et al. 2018). The central O abundances of \texttt{WD}s are results of stellar structure and evolution. Therefore, the asteroseismological research of \texttt{WD}s based on the \texttt{WDEC} (2018 version) procedure has an opportunity to explore the physical properties of progenitor stars. More asteroseismological work on \texttt{WD}s is required.

The optimal model of asteroseismology has a very small absolute difference of $\Phi$ = 0.06\,s. The stellar parameters of our optimal model are a little modifications to that of the previous spectral and asteroseismological work. The L19-2 is a massive and hot DAV star with relatively a thick H atmosphere and a thick He layer. In addition, for L19-2, the distance calculated according to our optimal model is only 1\% different from the distance reported by the latest Gaia Data Release 2. The pulsation mode of 118\,s is a trapped mode. For our optimal model, the rates of period change for the modes of 192\,s, and 113\,s are slightly larger but basically consistent with that of C$\acute{o}$rsico et al. (2016) when fitting the observed values. Our optimal model with the smallest $\Phi$ has an opportunity to reduce the uncertainty from the asteroseismological model and the period-fit procedure when calculating the theoretical $\dot{P}_{\rm cal}$. The smaller part that the theoretical value is smaller than the observed value has an opportunity to limit the mass of the axion.

\acknowledgements

This work is supported by the National Natural Science Foundation of China (Grant No. 11803004). We thank the supporting of Yunnan Province Youth Talent Project (2019-182) and the Academician Wang Jingxiu Workstation of Yunnan Province (202005AF150025).

\bibliographystyle{apj}


\end{document}